\RequirePackage{fix-cm}
\documentclass[smallextended]{svjour3}       
\smartqed  
\usepackage{amsmath}
\usepackage[dvips]{graphicx}
\usepackage{cite}
\usepackage{color,soul}
\usepackage{hyperref}
\journalname{Evolving Systems}
\begin{document}

\title{Convolutive Audio Source Separation using Robust ICA and an intelligent evolving permutation ambiguity solution
}


\author{Dimitrios Mallis         \and
        Thomas Sgouros  \and
        Nikolaos Mitianoudis
}


\institute{Dimitrios Mallis\at
              Electrical and Computer Engineering Dep.\\Democritus University of Thrace\\
			Xanthi 67100, Greece, \\
              \email{malldimi1@gmail.com}           
           \and
           Thomas Sgouros \at
               Electrical and Computer Engineering Dep.\\Democritus University of Thrace\\
			Xanthi 67100, Greece, \\
              \email{tsgouros@ee.duth.gr}
                \and
          Nikolaos Mitianoudis \at
               Electrical and Computer Engineering Dep.\\Democritus University of Thrace\\
			Xanthi 67100, Greece, \\
              Tel.: +30 25410 79572\\
              Fax: +30 25410 79569\\
              \email{nmitiano@ee.duth.gr}
}

\date{Received: date / Accepted: date}

\maketitle

\begin{abstract}
Audio source separation is the task of isolating sound sources that are active simultaneously in a room captured by a set of microphones. Convolutive audio source separation of equal number of sources and microphones has a number of shortcomings including the complexity of frequency-domain ICA, the permutation ambiguity and the problem's scalabity with increasing number of sensors. In this paper, the authors propose a multiple-microphone audio source separation algorithm based on a previous work of Mitianoudis and Davies\cite{mitianoudisDavies}. Complex FastICA is substituted by Robust ICA increasing robustness and performance. Permutation ambiguity is solved using two methodologies. The first is using the Likelihood Ration Jump solution, which is now modified to decrease computational complexity in the case of multiple microphones. The application of the MuSIC algorithm, as a preprocessing step to the previous solution, forms a second methodology with promising results.
\keywords{Convolutive Source Separation \and Robust ICA \and Permutation ambiguity}
\end{abstract}

\section{Introduction}

The problem of Blind Audio Source Separation (BASS) implies the extraction of independent audio sources from an audio mixture that has been observed by a number of microphones, without any prior knowledge regarding the involved sources or the mixing system. In recent years, many methods have been proposed for resolving this problem with relative success. BASS becomes more complicated, when we are dealing with real-room audio recordings. In reverberant rooms, each source is recorded multiple times by each microphone under different time delays and amplifications, due to sound waves' reflections on the room surfaces. The mixing system can thus be modelled using a room impulse response of finite length (FIR filter). In the general case of $M$ microphones that capture a mixture of $N$ sources, a common representation of the aforementioned mixing is $\boldsymbol{x}(t)=\boldsymbol{A} \ast \boldsymbol{s}(t)$, where $\ast$ denotes linear convolution, $ \boldsymbol{s}(t)=[s_1(t), s_2(t),..., s_N(t)]^T$ are the source signals, $ \boldsymbol{x}(t)=[x_1(t), x_2(t),..., x_M(t)]^T$ are the observation signals, $t$ is the time index and $\boldsymbol A$ is a matrix, whose elements $ \boldsymbol{a}_{ij}$ are FIR filters, describing the room impulse responses between the $j$-th source and the $i$-th microphone.

A classic decomposition method for performing BASS is {\it Independent Component Analysis} ({ICA})\cite{Comon94}. ICA extracts Independent Components ({ICs}) from a linear, instantaneous mixture, assuming independence between the original sources \cite{ICAbook01,ICAbook02}. In real rooms, where we deal with convolutive mixtures, ICA can also be applied by moving the separation to the frequency domain, where the convolution between the sources and the room transfer function is reduced to multiplication \cite{seki1998blind} for a number of discrete frequency bins $L$ i.e. $\boldsymbol{x}(f,t)=\boldsymbol{A}_f \boldsymbol{s}(f,t)$, where $f=1,\dots, L$. In other words, we transform a convolutional problem to a number of instantaneous problems, that can be solved using ICA.

By solving the separation problem in the frequency domain, ICA introduces 2 ambiguities: {\it scale} and {\it permutation}. The first results into random scaling of the extracted ICs, which can cause spectral deformations and reduce separation quality. The latter results into arbitrary source permutations along the discrete frequency bins, which consequently inhibits separation. The scale ambiguity can be resolved easily, as a post-processing step, by mapping the estimated sources back to the microphones' domain and recover the signals as they have been originally observed by the microphones {\cite{mitianoudisDavies,Mitiaphd}}.

The permutation ambiguity, on the other hand, is a difficult problem, and various techniques have been proposed without featuring robust performance in all cases. A weak coupling of neighboring frequency bins was proposed by Smaragdis \cite{seki1998blind}. Parra and Spence~\cite{Parra00IEEE} imposed a constraint on the unmixing filter length to be {\it smooth}, as they are modeled as FIR filters. Again mixed success has been reported for this method.  In \cite{ikeda1998method}, Ikeda and Murata proposed a time-dependent source envelope in order to align the permutations of the sources after separation. Mitianoudis and Davies \cite{mitianoudisDavies} also introduced a time-dependent source envelope, assuming Laplacian source priors, that is firstly introduced into the learning ICA algorithm to couple frequency bins during adaptation and secondly as a post-processing step using a Likelihood Ratio Jump to align permutations after the separation-learning procedure. A first attempt to use beamforming theory and direction of arrival information to align the permutations was introduced by Sawada et al. \cite{Sawada03}. Mitianoudis and Davies \cite{Mitia03d,Mitia04e} also examined the use of beamforming for permutation alignment with relative success. In \cite{Mazur09}, Mazur and Mertins align permutations by using generalised Gaussian Distribution in order to find differences between neighbouring frequency bins. Sawada et al. \cite{Sawada11} exploit the correlation coefficients of amplitude envelopes, which if maximised show the correct source alignment. Saito et al. \cite{Saito15} utilise, for the same purpose, the correlation between interfrequency power ratios. A different approach was followed by Sarmiento et al. \cite{Sarmiento15}, who find the spectral similarities between the separated components in the frequency domain, by employing a contrast function. A region-growing approach, to minimise the spreading of possible misalignments, in order to improve permutation alignment, was introduced by Wang et al. \cite{Wang11}. {Finally, Zhang and Chan \cite{Zhang10} proposed the minimal filter distortion (MFD) principle to overcome the separating filter indeterminacy in the separated sources. This is implemented by a least linear reconstruction error constraint of the separation system, which minimises the separating filter distortion.}

Most of the available methods for tackling the convolutive source separation problem are focused on the two-source two-microphone ($2\times 2$) case. However, low-cost commercially available hardware, such as the Microsoft Kinect interface, has been developed to offer low-latency four-microphone recordings and can be used to process $4\times 4$ cases. In this paper, we focus on the problem of audio source separation for determined cases (equal number of microphones and sources) that involves more than two sources. The presented methodology offers a computationally efficient solution for both the separation task as well as the permutation ambiguity. Based on the Kinect interface, we created a set of recordings containing mixtures of multiple sources as well as the original sources for evaluation purposes. This dataset is publicly available for further evaluation of audio separation methods\footnote{Dataset available at \texttt{http://utopia.duth.gr/nmitiano/download.html} }.  In this dataset, we will apply a novel framework that is optimized for multiple sources. This is then compared with the previous work of Mitianoudis and Davies \cite{mitianoudisDavies} to observe its efficiency for multiple sources. The proposed framework includes a robust complex ICA separation algorithm, called RobustICA \cite{zarozoRobustICA}, that has not been used before for convolutive audio source separation. In addition, we  present a new technique to tackle the permutation ambiguity problem, especially for a large number of sources, based on the Likelihood Ratio Jump solution \cite{mitianoudisDavies}. We show that this new technique (Reduced Likelihood Ratio Jump) can reduce the computational cost of addressing the permutation problem in comparison to the original Likehood Ratio Jump and can produce the same, if not better separation quality. Finally, the use of a beamforming algorithm, such as MuSIC, as a pre-processing step for the new Reduced Likelihood Ratio Jump method, for the lower frequencies, is investigated here. It appears that beamforming can expedite the convergence of the new approach, provided that the source separation problem falls within the requirements of a beamforming approach.  {Recently, Markovich-Golan et al \cite{Golan17} used the linearly constrained minimum variance (LCMV) beamformer for extracting individual sources in a mixture along with the “Triple N” ICA for convolutive mixtures (TRINICON) framework to estimate the room transfer functions between sources and microphones. A very interesting approach, where more microphones capture less sources.}

The paper is organised as follows. In the next section, we present the basics of instantaneous source separation, where the RobustICA method is discussed in more detail. Section 3 discusses convolutive source separation. In this section, the permutation problem is discussed, along with the previous and novel solutions we present here. Section 4 contains extensive evaluation of the proposed solutions. Finally, Section 5 concludes the paper.

\section{Instantaneous Complex Source Separation}
In the instantaneous case, we consider the following mixing process: $\boldsymbol{x}(t)=\boldsymbol{A}\cdot\boldsymbol{s}(t)$. In order to separate the sources, we have to estimate an unmixing matrix $\boldsymbol W$, such that $\boldsymbol{u}(t)=\boldsymbol{W}\cdot\boldsymbol{x}(t) \approx\boldsymbol{s}(t)$.

A common methodology to solve these kind of problems is {\it Independent Component Analysis} (ICA). This method estimates unmixing filters $w_{ij}$ that are arranged in a matrix format $\boldsymbol W$, assuming statistical independence between the source signals $s_1(t), s_2(t),..., s_N(t)$. The whole unmixing structure of ICA algorithms resembles the structure of a single-layer feed-forward neural network, whose weights $w_{ij}$ are updated by a backpropagation update rule. This update rule is derived by optimizing a cost function that emphasizes the statistical independence of the separated outputs. The rule is iterated over the whole training dataset in batch mode until the weights converge. Thus, ICA can also be considered a machine learning technique that can evolve over time to update the separation weights, i.e. an evolving system.

\subsection{The FastICA algorithm}

In the determined case, where the number of sources is equal to the number of observations ($N=M$), the most popular method of estimating the unmixing matrix $\boldsymbol{W}=[  \boldsymbol{ w_1 } , \boldsymbol{w_2} ,..., \boldsymbol{w_N} ]^T$ is the FastICA algorithm. There are many implementations of the FastICA algorithm, that are based on an optimization of a contrast function emphasizing nonGaussianity using a fixed-point iteration algorithm. One common fixed point algorithm is the following \cite{Hyva99NPL},
{
\begin{equation}
\Delta\boldsymbol{W}=\boldsymbol{D}[\textrm{diag}(-\alpha_i)+\mathcal{E}\lbrace \phi(\boldsymbol{u}) \boldsymbol{u^H}\rbrace]\boldsymbol{W}
\end{equation}
where $\phi(u)=u/|u|$ is an activation function for superGaussian sources, $\alpha_i$ an adaptive parameter, $^H$ is the Hermitian operator (complex conjugate transpose)  and $\mathcal{E}\{\cdot\}$ denotes the expectation operator. This iterative update of the unmixing matrix $\Delta\boldsymbol{W}$} is calculated using a Maximum Likelihood Estimator assuming Laplacian priors for the independent sources. The method also demands that the data of every frequency bin are prewhitenned. Whitening is the procedure of decorrelating the data via Principal Component Analysis (PCA) and normalising the decorrelated data to unit variance \cite{Hyva00NN}. Even though, this method has been initially derived for real-valued data mixtures, it has been adapted and has shown to work well with complex data in \cite{mitianoudisDavies}.

\subsection{The RobustICA algorithm}
In this section, we examine a source separation algorithm, named RobustICA \cite{zarozoRobustICA}. RobustICA optimizes the following generalized form of kurtosis.
{
\begin{equation}
\mathcal{K}(\boldsymbol{w})=\frac{\mathcal{E}\{|u|^4\}-2\mathcal{E}^2\{ |u|^2\}-|\mathcal{E}\{ u^2\}|^2}{\mathcal{E}^2\{ |u| ^2\}}
\end{equation}
where $\boldsymbol{w}$ denotes an unmixing vector and $u$ a separated source.} The above definition of kurtosis can be applied to both real and complex data. In addition, prewhitening is not necessary for RobustICA. RobustICA uses exact line search optimization of the absolute kurtosis contrast function, instead of fixed-point optimization, used by FastICA \cite{Hyva99TNN}.

\begin{equation}
\mu _{opt}=\arg \max _\mu (\mathcal{K}(\boldsymbol{w}+\mu \boldsymbol{g}))
\end{equation}
The search direction can be given by the gradient of kurtosis $\boldsymbol{g}= \boldsymbol{\nabla}_{\boldsymbol{w}}\mathcal{K}(\boldsymbol{w})$.
Exact line search is often a computationally expensive optimization technique that requires additional numerical analysis algorithms. In the case of kurtosis, the optimal step size $\mu _{opt}$ is calculated algebraically with a minimum computational cost. It is shown in \cite{zarozoRobustICA} that $\mu _{opt}$ can be calculated from the roots of a low-degree polynomial that maximizes the absolute value of the contrast function along the search direction. RobustICA has a number of advantages \cite{zarozoRobustICA} compared to the original FastICA:
\begin{itemize}
\item RobustICA does not make any assumption regarding the sources' statistical profile, and can deal with real and complex sources alike.
\item Prewhitening is not mandatory before RobustICA. Multiple ICs in that case can be extracted with the method of linear regression in contrast to symmetric orthogonalization that is used by FastICA.
\item The method can target sub-Gaussian or super-Gaussian sources in a specific order. This feature is useful in the audio separation case, where we know in advance that data in the frequency domain can be mostly modelled as super-Gaussian\cite{mitianoudisDavies}.
\item The method is robust to the presence of saddle points and spurious local extrema of the contrast function \cite{zarozoRobustICA}.
\item RobustICA can achieve great separation performance with relatively small additional computational cost, compared to other ICA implementations. This feature is demonstrated in \cite{zarozoRobustICA} and will be verified by the experimental results in this paper.
\end{itemize}

Despite the fact that prewhitening is not mandatory for RobustICA, it will be used as a preprocessing step in our proposed framework. This is due to the observation that it leads to a more computationally efficient implementation in the case of multiple sources. Since the prewhitened components lay on an orthogonal structure, every ICA iteration that attracts one IC towards an original source, forces the rest of the ICs to converge faster to  other sources.  This can be achieved with the use of symmetric orthogonalization, as in (\ref{eq:symmertic}).
{
\begin{equation}
\boldsymbol{W}^+\leftarrow\boldsymbol{W}(\boldsymbol{W}^H\boldsymbol{W})^{-0.5}
\label{eq:symmertic}
\end{equation}}

On the contrary, in linear regression, after the extraction of an IC, we have to separate a reduced mixture from a random position, which can be rather slow. As a result, we use prewhitening to improve the convergence speed of our method, in expense of the separation performance limitations that prewhitening can introduce.

\section{Frequency-domain source separation}
Frequency-domain source separation methods apply the Short-Time Fourier Transform (STFT) to the mixture recordings $\boldsymbol{x}(t)$. Consequently, the convolutive mixture is transformed to $L$ instantaneous mixture via the STFT, i.e. $\boldsymbol{x}(t)=\boldsymbol{A} \ast \boldsymbol{s}(t)\Rightarrow \boldsymbol{X}(f,t) = \boldsymbol{A}_f \boldsymbol{S}(f,t)$,  {where $f$ denotes the discrete frequency bin, $t$ the discrete time index of the STFT.} The separation problem can be solved independently using any complex ICA algorithm, such as RobustICA. {That is to say, we aim at estimating a complex unmixing matrix $\boldsymbol{W}_f=\boldsymbol{A}_f^{-1}$ at each frequency bin $f$ that will separate the sources, again at each frequency bin $f$,  i.e. $\boldsymbol{u}(f,t) = \boldsymbol{W}_f \boldsymbol{X}(f,t)$}. ICA's inherent scale and permutation ambiguities impose severe problems in this framework and must be resolved. Scale ambiguity is tackled using a mapping to the microphone domain\cite{mitianoudisDavies}. There exist many methods to tackle the permutation ambiguity of frequency-domain BASS methods, as mentioned in the introduction.
\subsection{Likelihood Ratio Jump}
Mitianoudis and Davies introduced the Likelihood Ratio Jump method in \cite{mitianoudisDavies} for the alignment of frequency bins to the correct source.  This method can be used either after each iteration of the ICA algorithm, or even better as a post-processing mechanism. The method works iteratively and in each iteration it forms a {\it likelihood ratio jump} to decide, which permutation is the most probable for each frequency bin. It uses a set of rescaling parameters $\gamma_{ij}$ that model the probability of the $i^{th}$ source moving to the $j^{th}$ position. For each frequency bin, it calculates the probabilities for all possible permutations. For example, in a mixing of 3 sources a possible permutation of the extracted ICs: $IC 3 \rightarrow IC 1 $, $IC 1 \rightarrow IC 2$, $IC 2 \rightarrow IC 3$, forms the probability:
\begin{equation}
L=-\log(\gamma_{31}\gamma_{12}\gamma_{23}) \label{eq:loglikelihood}
\end{equation}

The correct permutation is the one that produces the maximum probability,  as in (\ref{eq:loglikelihood}). For the case of three sources, there are $3!=6$ possible permutations for the extracted ICs, which have to be assessed probabilistically, to conclude which permutation is the most likely to be correct. The parameters $\gamma_{ij}$ are produced through a maximum likelihood estimator and can be calculated as follows:
\begin{equation}
\gamma_{ij}=\frac{1}{T}\sum_t{\frac{|u_i(f,t)|}{\beta_j(t)}}
\end{equation}
where $u_i(f,t)$ is the value of $i$-th IC for the discrete frequency bin $f$ and time index $t$ and { $\beta_j(t)$ is a non-stationary time-varying scale parameter (time envelope)} that is calculated for the source $j$. Finally, $T$ is the number of {distinct time frames of the STFT}.

The parameter $\beta_j(t)$ incorporates information related to the signal's spectral envelope over time, thus it can be interpreted as a volume measurement {along time}. Literally, it measures the overall signal amplitude along the frequency axis, emphasizing the fact that one source is ``louder'' than others at a certain time slot. This ``temporal energy burst'' can force the alignment of the permutations along the frequency axis. A possible estimation for the $\beta_j$ parameter can be the following:
\begin{equation}
\beta_j(t)=\frac{1}{L}\sum_f{|u_j(f,t)|}
\end{equation}
where $L$ is the number of frequency bins. Since the audio signal representation is sparse in the frequency domain\cite{mitianoudisDavies}, we assume a Laplacian source prior and thus the ``energy'' measurement in the estimation of $\beta_j(t)$ follows the L1-norm, instead of the commonly used L2-norm. The {\it Likelihood Ratio Jump} (LRJ) method has demonstrated very stable performance in solving the permutation ambiguity for a large number of cases \cite{mitianoudisDavies}. However, this was mainly demonstrated for $2\times 2$ cases.

{The parameter $\beta_j(t)$ is evolving over each iteration of the algorithm from a time envelope of the mixed signal to the time envelope of the source $j$. It is self-adapting using a sparse Bayesian model to aim for sparse sources, as mentioned earlier, which gives its adaptation to the unknown and unpredictable structures of the input audio sources, that can not be predicted or specified from the start. In essence, it is a Bayesian evolving system that adapts to the model of the input audio sources. In \cite{Herbig11}, Herbig et al proposed an adaptive scheme for a speech recognition system so as to adapt to new speakers. This is similar to the parameter $\beta_j(t)$, which essentially adapts to each source signal, in order to sort out the permutation ambiguity.}

\subsection{Reduced Likelihood Ratio Jump (RLRJ)}
One major disadvantage of this method is its computational cost that increases rapidly with the number of sources, as for each iteration of the algorithm we need to make $N!$ comparisons. For example, we can consider a case of $5$ sources, where the FFT has $4096$ frequency bins, and the post-processing permutation sorting method needs to spend $15$ iterations for the system to converge to the correct permutation for most of the bins. In total, we will need $5! \times 2,048 \times 15 =120 \times 2,048 \times 15=3,686,400$ calculations of the expression in \eqref{eq:loglikelihood} and as a result the whole task is computationally inefficient, if not prohibitive.

In this section, we propose a new ``suboptimal'' method, named {\it Reduced Likelihood Ratio Jump} (RLRJ). This technique selects to perform a few major comparisons, in contrast to the full set of $N!$ comparisons in the original method, thus the term ``suboptimal''. Nonetheless, we witnessed that it can produce the same, if not better separation quality with a considerable reduction of the computational cost.

RLRJ is based on the iterative nature of the original method. The original LRJ needs, in most of the examined examples, some dozen iterations for every frequency bin to converge to the correct permutation. This is mainly due to the parameter $\beta_j(t)$. As previously mentioned, this parameter incorporates information about the time envelope of the signal. As more permutations are sorted in each iteration, the time envelope of each signal becomes more distinct and as a result, the parameter $\beta_j(t)$ has a stronger impact in the calculation, that helps resolving the permutation for frequency bins, where this task is more difficult.

During extensive experimentation, we witnessed that there are many frequency bins that feature the correct source permutation from the first or second iteration of the method. For the remaining frequency bins, the algorithm after some iterations needs to swap only one IC pair to restore the correct permutation, since the rest have already been sorted to the correct sources. This situation is common for cases with many sources, as the correct permutation is clear for most of the ICs after a small number of iterations, and only one or two pairs may need an improved calculation of the parameter $\beta_j(t)$ to be permuted correctly.

Based on the above observation, we propose to reduce the number of examined permutations in every iteration of the algorithm. As the method works iteratively, we propose to calculate the most probable permutation from a set of permutations that includes the swap of only one pair of ICs at a time. In the case of $N$ sources,
we can calculate only the most probable from the $N-1$ swaps between the ICs at every iteration. As the method evolves over time, only one swap will be needed to ensure the correct permutation of the sources. Even if more than one pairs of ICs are permuted incorrectly, as the method progresses, the correct permutation will be restored, one pair at a time. By examining only one pair of permutations at a time, we reduce the complexity of the method from $N!$ to $\frac{1}{2}N(N-1)+1$, which for 5 sources means a reduction from 120 to 11 comparisons per iteration per frequency bin. {There is no theoretical bound to the number of $N$ sensors/sources that will make permutation sorting impossible. There are only practical limitations, including the number of available microphones and the available DSP's processing power, if we need to perform separation in real-time.}

\begin{figure}
\centering
\includegraphics[width=.8\textwidth]{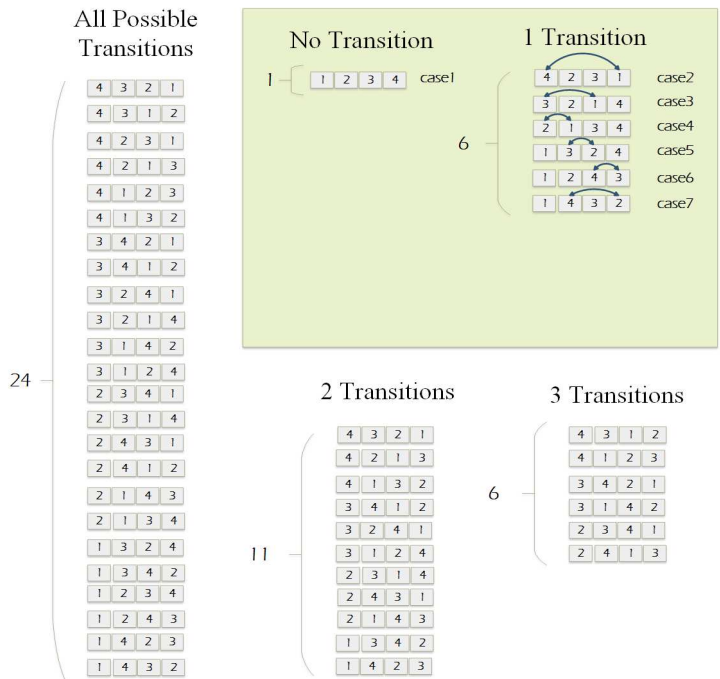}
\caption{All possible comparisons the LRJ needs to make for 4 sources, if we allow maximum 1, 2, 3, 4 transitions.}
\label{fig:figure0-5}
\end{figure}

Most importantly, as we will show in the experimental section, this suboptimal method does not undermine the quality of the separation but may also enhance it instead. An example of all possible transitions for $4$ sources is depicted in Fig. \ref{fig:figure0-5}. There, we can see the number of comparisons that are needed to be made, if we allow maximum 4, 3, 2 and finally 1 transition.

\begin{figure}
\centering
\includegraphics[width=.8\textwidth]{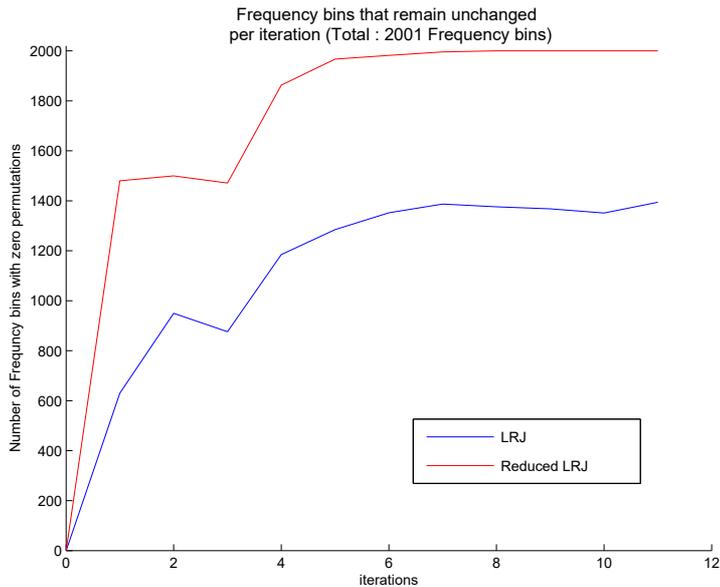}
\caption{Comparison between Reduced LRJ and original LRJ. More frequency bins remained unchanged using the Reduced LRJ for the same number of iterations.}
\label{fig:figure-RLRJ-LRJ}
\end{figure}

In Fig.~\ref{fig:figure-RLRJ-LRJ}, we can see that the Reduced Likelihood Ratio Jump converges for more frequency bins, compared to the original method for the same number of iterations. In an example of a total of 2001 frequency bins the Reduced LRJ has concluded the permutation sorting of all the frequency bins from the 8th iteration, in contrast to the original method that always needs to sort about 600 more frequency bins. For some bins, the original LRJ changes permutation in every iteration. This phenomenon is due to the very small differences between the likelihood values that force the original method to toggle between 2 permutations for specific frequency bins unnecessarily.

\subsection{RLRJ initialisation using Beamforming (MuSIC-RLRJ) }
The efficiency and accuracy of the previous permutation sorting framework, depends highly on the $\beta_j(t)$ parameter. We observed that as the estimation of $\beta_j(t)$ improved over several iterations, the RLRJ was able to determine the correct permutation for the frequency bins. Early inaccurate estimates of $\beta_j(t)$ made the task of aligning permutations harder. Based on this observation, we will attempt to enhance our framework by starting the RLRJ with a more accurate estimate of $\beta_j(t)$ compared to the one induced by the source separating algorithm. To accomplish that we will perform a source permutation sorting along frequency by incorporating information related to the geometry of the auditory scene \cite{JimGriffiths}. This step, apart from solving the permutation problem for a number of frequency bins, will enable us to start the RLRJ with a better estimate of $\beta_j(t)$, making the proposed framework more robust.

In order to incorporate the information, regarding the source positioning in the auditory scene, we will use  {\it Beamforming}. Beamforming can manipulate the overall gain of a sensor array of known geometry, in order to focus on a desired source, that comes from a specific direction, while suppressing other sources from different directions. In the general case, beamforming assumes the existence of more sensors than observed source signals, placed on different angles related to the centre of the sensor array.

Beamforming can be used to estimate the angles $\theta_i$ at which the source signals arrive at the origin of the array's coordinates system. These angles are called {\it Directions of Arrival} (DOA) for the sources in the far-field approximation. In the frequency-domain source separation scenario, we work independently for each frequency bin. Thus, we can model each frequency bin's content as narrow-band, i.e. $s_i(t)=\alpha e^{j2\pi f t}$ with a carrier frequency $f$, representing the frequency of each frequency bin. For the case of only one source, different sensors with distances $d_j$ from the array centre, capture the source signal $s_i$ with a time lag $T_j$, since the speed of sound is finite. The delays $T_j$ are functions of the signals DOA $\theta_i$. Most sensor arrays feature equal distance between the sensors. This implies that the delay of the $j$-th sensor will be given by $T_j=(j-1)T$, where $T$ is the signal delay between the first two sensors, assuming that the first sensor is the centre of the array. Based on this analysis, we can produce the following model,

\begin{equation}
\boldsymbol{x}\approx
\begin{bmatrix}
1 \\
\alpha e^{-j2\pi f T} \\
...\\
\alpha e^{-j2\pi f (M-1)T} \\
\end{bmatrix}
s_1 (t)=\alpha(\theta_1)s_1(t)
\end{equation}
where $T=d \sin\theta_i /c$, $d$ is the distance between each sensor pair and $c=430m/s$ is the velocity of the air. A plot of $\alpha(\theta)$ is called {\it directivity pattern} of the array.

If we generalize for $N$ sources we have the system $\boldsymbol{x}=\boldsymbol{A}\boldsymbol{s}$, where $\boldsymbol{A}=[\alpha(\theta_1) \alpha(\theta_2) ... \alpha(\theta_N)]$. As we can see, the model of a linear sensor array is similar to the ICA framework model. The main difference is that this model incorporates more geometrical information (i.e. Directions of Arrival, position of sensors), whereas in source separation the mixing model is more general, employing sources statistical information only.

Beamforming's main objective is to estimate $\boldsymbol{A}$ and separate the audio sources. Despite its many applications in telecommunications, it is not exactly a very efficient tool for performing audio source separation. This is because beamforming assumes more sensors than sources, which does not agree with our requirements. {Even though, this seems prohibitive, there is already a work-around in our separation framework. Using an ICA algorithm (FastICA, RobustICA), we can separate all input sources at each frequency bin $f$. To solve the scale ambiguity, we map each source back to the microphone domain, i.e. we have $N$ signals for each source, one observation at each microphone. This implies that we have more sensors than sources after separation and scale ambiguity correction, thus, we can apply any beamforming approach \cite{Mitiaphd}.}

The other problem is that the audio sources that propagate in the room are quite wide-band. Since, we are solving beamforming independently at each frequency bin, there is a vast number of frequencies that need to be addressed. As shown by Mitianoudis in \cite{Mitiaphd}, for a given sensor distance $d$, there is a frequency bound $f\le c/2d$ above which  multiple nulls around the DOA appear in the calculated directivity patterns. This is known as {\it spatial aliasing}. As a result, the estimation of the real DOA becomes unclear for frequencies above this bound. Finally, the room's impulse response introduces small displacement of these DOA along frequency, due to the existence of multipath in the room.

Despite these disadvantages, we can achieve an accurate computation for the DOAs in the auditory scene, for cases where the sensor spacing $d$ leaves enough frequency range intact, while enabling sensors to capture the sources in a distinct manner and the room reverberation is not severe. In our case, we will use the four-microphone array that exists in Microsoft Kinect, where the sensor spacing is not equal. We will use the MuSIC algorithm to get an estimate of the DOA that exist in the scene and then we will align permutations of each frequency bin, according to the estimated DOA for all frequencies below $f\le c/2d$. This preprocessing step will improve the calculation of $\beta_j(t)$ and increase the performance of the RLRJ.

\subsubsection{The MuSIC Algorithm}

One of the methods that can be used for a very accurate estimation of DOA is the {\em Multiple Signal Classification} (MuSIC) algorithm \cite{JimGriffiths,Cheney01thelinear}. MuSIC is a subspace method based on decomposition of the observations covariance matrix. To formulate the MuSIC algorithm, we will incorporate a noise component $\boldsymbol{\epsilon}(t)$ to our original model.

\begin{equation}
\boldsymbol{x}(t)=\boldsymbol{A} \boldsymbol{s}(t) + \boldsymbol{\epsilon}(t)
\end{equation}

The covariance matrix of $\boldsymbol{x}$ can be calculated as follows:

\begin{equation}
C_x=\mathcal{E}\{\boldsymbol{x}\boldsymbol{x}^H \}=\boldsymbol{A}\mathcal{E}\{\boldsymbol{s}\boldsymbol{s}^H\}\boldsymbol{A}+\mathcal{E}\{\boldsymbol{\epsilon}\boldsymbol{\epsilon}^H\} \\
\end{equation}

\begin{equation}
C_x=\boldsymbol{A}C_s\boldsymbol{A}^H +{\sigma_\epsilon}^2 I
\end{equation}
where $C_x$ is the covariance matrix of $\boldsymbol{x}$, $C_s$ is the covariance matrix of $\boldsymbol{s}$ and $C_\epsilon={\sigma_\epsilon}^2 I$ the covariance matrix of noise that is considered additive and isotropic. It can be shown that $C_x$ has $M$ eigenvalues, $N$ of which correspond to the sources and $M-N$ to noise.

As shown in \cite{Moonbook}, the space spanned by the columns of matrix $\boldsymbol{A}$ is equal to the space spanned by the eigenvectors $ E_s =[\boldsymbol{e}_1,\boldsymbol{e}_2,...,\boldsymbol{e}_N]$.

\begin{equation}
\mathtt{span}\{\boldsymbol{A}\}=\mathtt{span}\{[\boldsymbol{e}_1,\boldsymbol{e}_2,...,\boldsymbol{e}_N]\}=\mathtt{span}\{E_s\}
\end{equation}
where $E_s=[\boldsymbol{e}_1,\boldsymbol{e}_2,...,\boldsymbol{e}_N]$ are the eigenvectors corresponding to the desired sources and $E_n=[\boldsymbol{e}_{N+1},\boldsymbol{e}_{N+2},...,\boldsymbol{e}_M]$ the eigenvectors corresponding to noise.

In practice, we can plot the following function

\begin{equation}
M(\theta)=\frac{1}{(\boldsymbol{\alpha}^H E_n {E_n}^H \boldsymbol{\alpha})^2} \forall\theta \in [-90^o,90^o]
\end{equation}

The N peaks of function $M(\theta)$ will denote the DOA of the N sources.

\subsubsection{Permutation sorting with Beamforming (MuSIC-RLRJ)}

In the previous subsection, we saw how we can acquire an estimation for the source DOA observed by multiple microphones using the MuSIC algorithm. In order to use this technique in our framework, we first have to overcome the requirement of more observations than sources. This can be achieved using the method that resolves the scale ambiguity. By mapping the estimated sources back to the microphones domain, we recover the signals as they have original been observed by the microphones. As a result, for each extracted source, we can have as many observations as the microphones of the Kinect sensor. Thus, we can apply the MuSIC method to derive a $M(\theta)$ function for every estimated source. The peak of $M(\theta)$ is the DOA of this source, related to the microphone array.

As mentioned earlier, the beamforming method is only going to resolve the permutations for the frequency bins with lower frequencies. As shown in \cite{Mitiaphd}, we can emphasize the position of the DOA for the whole signal by averaging the $M(\theta)$ plots for all frequency bins below 2KHz. The idea is that, although the permutations are not sorted, the average of all $M(\theta)$ plots will present picks around the real DOA of the original sources. As a result, the peaks of the following function will represent the real DOA.

\begin{equation}
P(\theta)=\sum_{f \in [0-2KHz]} \sum_{i=1}^N {M_i (f,\theta)}
\end{equation}

By plotting the function $P(\theta)$, we can estimate the direction of arrival for the observed sources. We have to mention that this approach can not always produce useful results. MuSIC does not perform well when the sources are highly correlated, and the auditory scene has to be clear. The corresponding sources must have some angular distance between them, so that the DOA can be distinct in the plot. Also in rooms with high reverb, the $P(\theta)$ plot presents multiple picks that correspond to reflections of the sound waves in the surfaces of the room. To be able to estimate the DOA from $P(\theta)$ function, the following restrictions have to be fulfilled.

\begin{itemize}
\item The minimum angular distance between sources that can be observed has to be about 15 degrees.
\item There must be at least as many peaks as the number of sources in $P(\theta)$. In a different case, some sources may have been integrated under the same DOA and the method can not be used.
\end{itemize}

To solve the permutation ambiguity for the lower frequency bins, $P(\theta)$ is divided into distinct regions. Each region corresponds to each source. At each frequency bin, we then permute the sources, so that each source features a DOA inside the corresponding source DOA region.

\begin{figure}
\centering
\includegraphics[width=.8\textwidth]{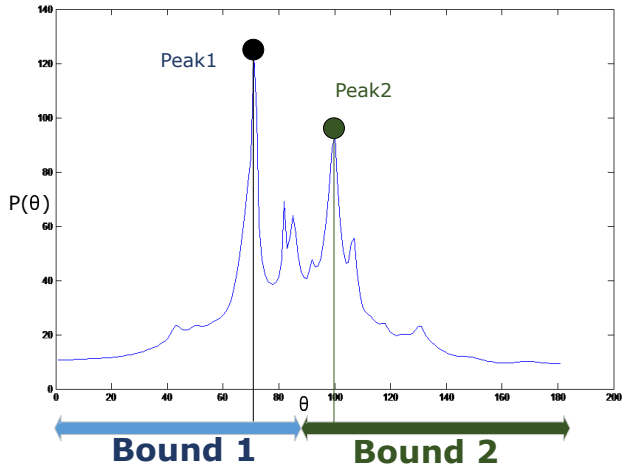}
\caption{DOA estimation using $P(\theta)$ for a 2 source - 2 microphones real room recording.}
\label{fig:figureDOAest}
\end{figure}

In Fig. \ref{fig:figureDOAest}, we can see an example of an aggregated $P(\theta)$ plot for a real room recording of a 2 source-microphones case. As we can see, the diagram presents 2 clear peaks that denote the DOAs of the observed sources. Because the recording took place in a real reverberant room, we can observe also smaller peaks coming from different angles, which is due to the strong multi-path. As described earlier, we split the angles of arrival in 2 regions, and the permutation ambiguity is sorted accordingly for every frequency bin. The separated outputs are then presented to RLRJ for further processing.

\section{Experiments}

\subsection{Evaluation process}
\begin{figure}
\centering
\includegraphics[width=\textwidth]{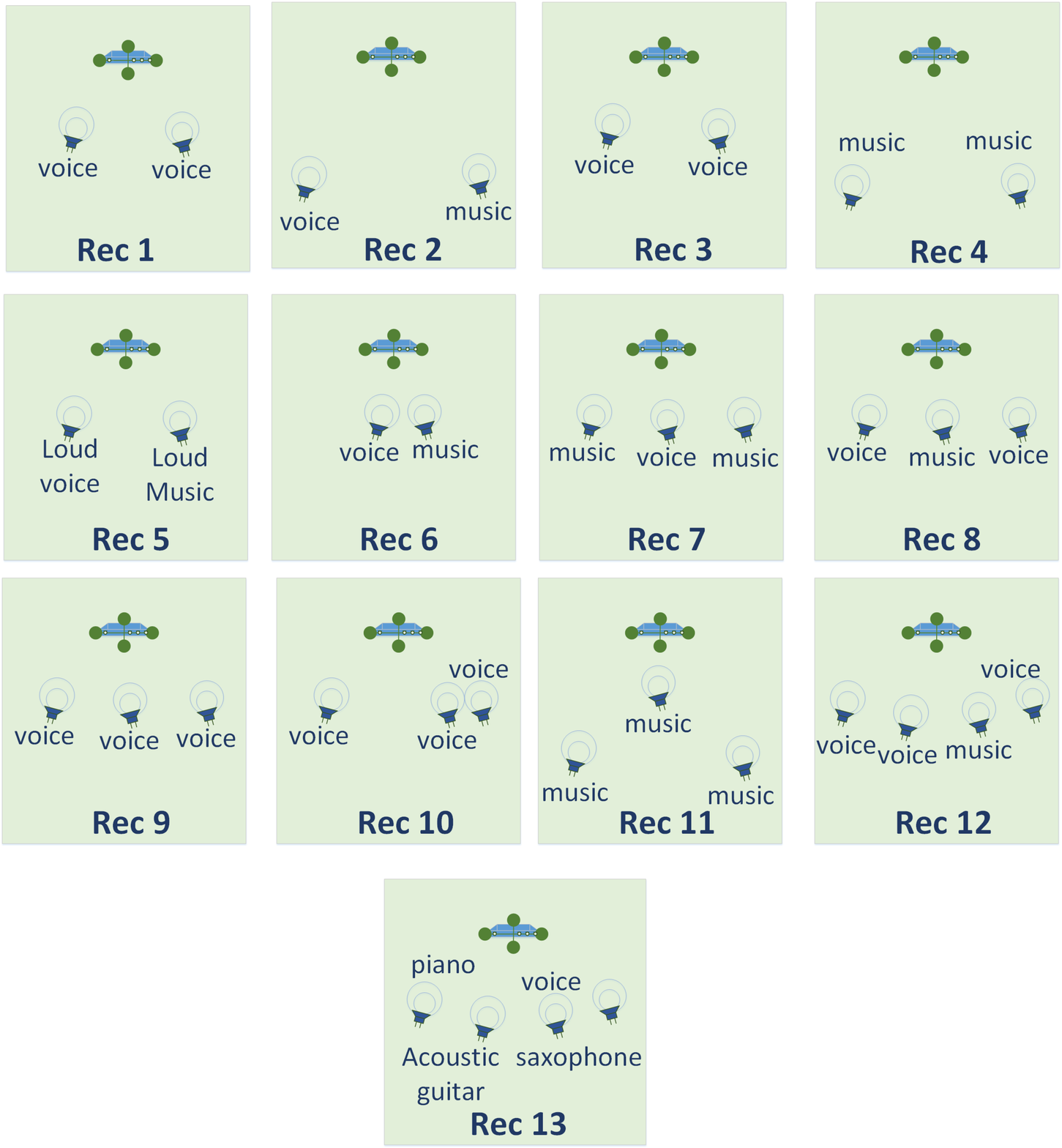}
\caption{The different sources-microphones positions that are included in the Source Separation Dataset.}
\label{fig:figureRecPos}
\end{figure}

To evaluate the performance of our proposed framework, we created an evaluation dataset of $13$ audio recordings. This evaluation dataset contains $6$ mixtures of $2$ sources, $5$ mixtures of $3$ sources and $2$ mixtures of $4$ sources, featuring both speech and music. The recordings were made using the Microsoft Kinect interface. The sources-microphones were placed in different positions in a real reverberant room.  Fig. \ref{fig:figureRecPos} depicts the different source-microphone positions that were recorded. It also includes solo recordings of the corresponding sources under the same recording conditions (position and loudness) in order to be used as ground truth for the evaluation of the separation quality, achieved by the separation framework. This dataset is publicly available for download\footnote{Dataset available from \url{http://utopia.duth.gr/nmitiano/download.html}}.

We used the RobustICA MATLAB implementation, as provided freely by the authors\footnote { \url{http://www.i3s.unice.fr/~zarzoso/robustica.html}}, and we made the necessary modifications so that it can work in a convolutive source separation framework using the LRJ,  RLRJ and MuSIC-RLRJ permutation sorting methods. For the experiments presented in this section, we assume a room impulse response of max 90.7 ms length (4096 samples at 44.1 KHz), which we reckon is a valid model for the recording positions in the reverberant room. { To measure the separation quality, we used the metrics {\em Signal-to-Distortion Ratio} (SDR), the {\em Signal-to-Interference Ratio} (SIR) and the {\em Signal-to-Artifact Ratio} from the BSS$\_$EVAL Toolbox v.3~\cite{vinc}. These metrics are measured in dB, which implies that higher values denote better performance and there is no upper bound. They are designed specifically for Performance Measurement in Blind Audio Source Separation. SDR compares the separated with the corresponding original source and calculates the ratio of the source signal energy versus the energy of interfence from other sources, separation artifacts and possible noise. SIR calculates the ratio of the source signal energy versus the energy of interference from other sources. Finally, SAR calculates the ratio of the source signal, interference and noise energy versus the artifacts energy.} These metrics are implemented as a publicly-available MATLAB Toolbox and were designed to allow a time-invariant filter distortion of $512$ samples length. We manually changed this value to $2000$ samples, to cater for the incorrect synchronization between the original and the estimated sources that are extracted from the mixture \cite{vinc}.

\begin{figure}
\centering
\includegraphics[width=\textwidth]{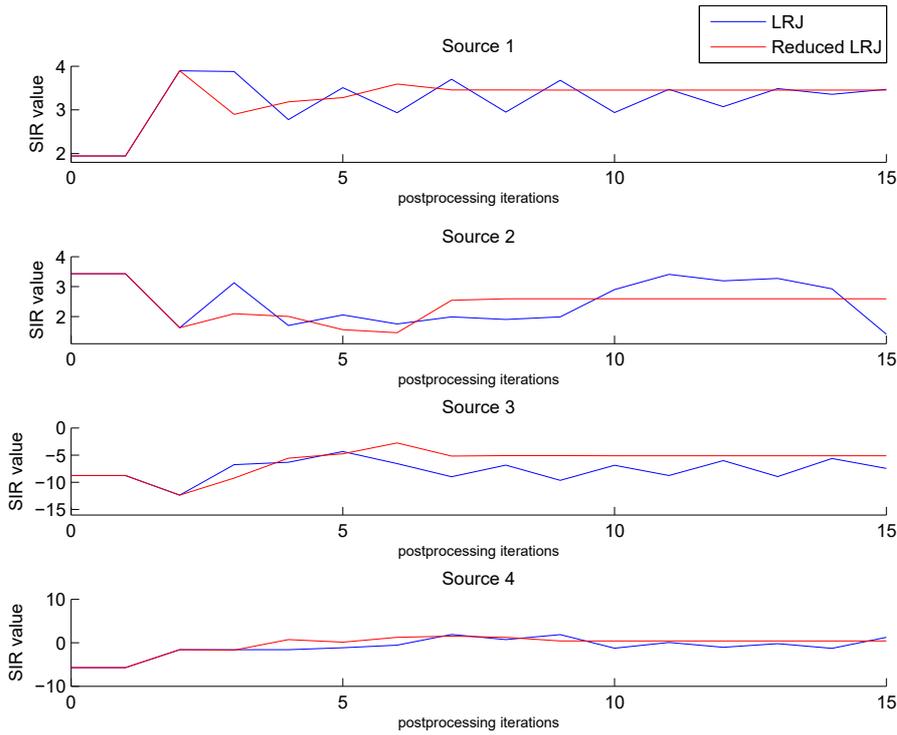}
\caption{Separation Quality in terms of SIR for the LRJ and the RLRJ methods.}
\label{fig:figureRJ-4sources}
\end{figure}

Metrics SDR and SAR measure the distortion and artifacts that are created by the separation method. Both examined frameworks (the proposed framework and Mitianoudis-Davies) produce similar values as they employ similar contrast functions with different optimization methods. To avoid the repetition of similar experimental results, we will therefore present SIR measurements only, and the separation quality in terms of interference elimination between the extracted sources.

\begin{table}
\renewcommand{\arraystretch}{1.3}

\caption{Efficiency Comparison between the Frequency Domain RobustICA and FastICA implementations in terms of Signal-Interference-Ratio (SIR) (dB). Here, we compare the performance of 3 and 15 iterations of RobustICA and FastICA. RobustICA reaches a better separation result faster than FastICA.}
\label{table2}
\centering
\begin{tabular}{cccccc}
\hline
\bfseries Recording & \bfseries Source & \multicolumn{2}{c}{3 iterations} & \multicolumn{2}{c}{15 iterations}  \\
\hline
 &  & RobustICA  & FastICA & RobustICA & FastICA  \\
 \hline
 & 1 & -0.464 &  \textbf{-0.09} & -0.09 & \textbf{0.88} \\
7 & 2 & \textbf{1.698} &  0.99 & 0.60  & \textbf{1.51}\\
 & 3 & \textbf{-7.30} &  -7.98 & -8.84 & \textbf{-7.99}\\
\hline
 & 1 & -6.36 &  \textbf{-4.25} & -6.79 & \textbf{-3.37}\\
8 & 2 & \textbf{5.53} &  0.23 &\textbf{ 3.46} & 1.07 \\
 & 3 & \textbf{1.12} &   -2.00 & \textbf{0.67} & -0.38\\
\hline
 & 1 & \textbf{5.40} & 3.20 & \textbf{6.44} & 4.20\\
9 & 2 & \textbf{0.18} &  -1.17 & \textbf{-0.63} & -2.38\\
 & 3 &  -10.11  & \textbf{-8.98} & -10.31 & \textbf{-9.78}\\
\hline
 & 1 & -2.40  & \textbf{-2.16} &  -3.34 &\textbf{ -0.10}\\
10 & 2 &  \textbf{2.26}  & 1.75 &  2.82 & \textbf{3.53} \\
 & 3 & -6.78  & \textbf{-6.22} & \textbf{ -6.24} & -10.09\\
\hline
 & 1 & \textbf{3.81}  & -2.47 & \textbf{4.12} &   1.12\\
11 & 2 & \textbf{-5.78}  &  -6.05 & \textbf{-6.61} & -8.02\\
 & 3 &  \textbf{4.01}  & 0.99 & \textbf{4.37} & 3.05\\
\hline
 & 1 & -6.80  & \textbf{-1.33} & -6.79 &  \textbf{ -4.02}\\
12 & 2 & \textbf{-3.78}  &  -6.82 &  -4.57 & \textbf{-3.72}\\
 & 3 &  \textbf{1.18}  & -0.05 & 1.73 & \textbf{4.27}\\
  & 4 &  \textbf{-0.63}  & -2.17 & \textbf{-0.95} & -1.50\\
\hline
 & 1 & \textbf{2.44}  &  1.19 & 0.50 &  \textbf{ 0.64}\\
13 & 2 & -1.68  &  \textbf{0.30} &  \textbf{-0.25} & -0.77\\
 & 3 &  \textbf{-6.99}  & -7.68 & \textbf{-6.72} & -7.87\\
  & 4 &  -7.83  & \textbf{-3.38} &  -5.17 & \textbf{-5.02}\\
 \hline

\end{tabular}
\end{table}

\begin{table}
\renewcommand{\arraystretch}{1.3}

\caption{Comparison between the Likelihood Ratio Jump (LRJ) and the Reduced Likelihood Ratio Jump (Reduced LRJ) for 8 available iterations in terms of Signal-Interference-Ratio (SIR) (dB). The term rec refers to the recording id in the dataset.}
\label{table3}
\centering
\begin{tabular}{ccccccc}
\hline
Method & Source & \bfseries rec 7 & \bfseries rec 8 & \bfseries rec 9 & \bfseries rec 10 & \bfseries rec 11  \\
\hline

 & 1 & -0.24 & -6.46 & 6.18 & -1.55 & 3.18\\
LRJ & 2 &  1.58 & \textbf{5.42} & 0.29 & 1.13 & -7.64\\
 & 3 & \textbf{-6.98} &  \textbf{1.03} & -10.71 &  -6.49 & 2.41\\

\hline
 & 1 & \textbf{2.72} & \textbf{-2.82} & \textbf{6.44} & \textbf{-0.85} & \textbf{3.82}\\
Reduced LRJ & 2 &  \textbf{1.67} & 3.82 & \textbf{1.91} & \textbf{1.46} &\textbf{ -5.78}\\
 & 3 & -7.87 &  -2.57 & \textbf{-10.30} & \textbf{ -5.59} &\textbf{ 4.01}\\

\hline
\end{tabular}
\end{table}

\begin{table}
\renewcommand{\arraystretch}{1.3}
\caption{Running time comparison between the two frameworks (in seconds). Framework 1 refers to the new proposed framework (RobustICA + Reduced Likelihood Ration Jump). Framework 2 refers to the previous framework (FastICA+ Likelihood Ratio Jump). Separation Time refers to the running time of the source separation algorithm (RobustICA/FastICA). Permutation Time refers to the running time of the permutation ambiguity sorting time.}
\label{table4}
\centering
\begin{tabular}{ccccccc}
\hline
\bfseries Recording  & \multicolumn{3}{c}{Framework 1} & \multicolumn{3}{c}{Framework 2}  \\
\hline
  & Separation& Permutation & Total & Separation & Permutation & Total   \\
  & Time & Time & Time & Time  & Time & Time   \\
 \hline
7 (3 sources) & \textbf{5.58} &  \textbf{5.89} & \textbf{13.04} & 14.38 & 7.16 & 23.18\\
8 (3 sources) & \textbf{4.89} & \textbf{5.21} & \textbf{11.43} & 11.30 & 6.00 &  18.61\\
9 (3 sources) &  \textbf{4.45} & 2.71 & \textbf{7.70} & 5.27 & \textbf{2.56} & 8.39\\
10 (3 sources) & \textbf{4.63} & 2.96 & \textbf{8.20} & 5.92 & \textbf{2.92} & 9.48\\
11 (3 sources) &  \textbf{5.00} & \textbf{5.24} & \textbf{11.60} & 11.61 & 6.16 & 19.17\\
12 (4 sources) &  \textbf{6.73} & \textbf{8.93} & \textbf{17.36} & 22.25 & 17.86 & 41.82\\
13 (4 sources) & \textbf{6.99} & \textbf{10.26} & \textbf{19.27} & 22.68 & 20.58 & 45.29\\
\hline
\end{tabular}
\end{table}

\subsection{Performance Evaluation}

\subsubsection{Source Separation Algorithm Comparison}

In this section, we present several experiments to demonstrate the efficiency  of the 2 examined frameworks using FastICA, RobustICA, the LRJ and the RLRJ. Firstly, in Table \ref{table2} we compare the separation quality of the two ICA implementations in terms of SIR. To tackle the permutation ambiguity, we use, in this experiment, 5 iterations of the original LRJ of Mitianoudis and Davies \cite{mitianoudisDavies} for the two frameworks. We can see that:
\begin{itemize}
\item As the 2 methods perform optimization of different criteria, they do not perform the same for the examined cases. RobustICA performs better for cases (8,9,11), FastICA for (10,12) and for the rest of the recordings, we observe similar separation qualities. In general, we can say that for the examined cases, RobustICA with prewhitening can reach and outperform slightly the original method of FastICA in separation quality.
\item RobustICA presents very fast convergence. In all examined cases, it produces very good separation quality in only 3 iterations. This feature of RobustICA can be a great advantage, compared to previous ICA implementations, as also mentioned in \cite{zarozoRobustICA}. In contrast, FastICA needs more iterations to produce stable results. We can see, in Table \ref{table2}, the major differences in separation quality from 3 to 15 iterations of FastICA, in comparison to RobustICA that reaches very good separation quality in only 3 iterations, which is then only slightly improved  as the iterations increase. Despite the fact that a RobustICA iteration is more costly than the FastICA  equivalent, its fast convergence improves the total computational efficiency.
\end{itemize}

\subsubsection{Comparison between LRJ and RLRJ}
In the next experiment, we compare the separation quality with 8 iterations for both the new Reduced (RLJR) and the original Likelihood Ratio Jump (LRJ) method of Mitianoudis and Davies \cite{mitianoudisDavies}. For source separation in this experiment, we have used the RobustICA with prewhitening, as it has shown to be the most efficient method. In Table \ref{table3}, we can compare the separation performance of the two permutation solving methods for three-source recordings. The LRJ method performs better only in recording 8. For the rest of the recordings, the Reduced Likelihood Ratio performs better despite the fact that it is a suboptimal method.

This improved performance of RLRJ can be due to the stability that is produced from the convergence of a greater number of frequency bins, as shown previously in Fig. \ref{fig:figure-RLRJ-LRJ}. The Reduced Likelihood Ratio Jump, by allowing a smaller set of possible swaps, leads to a more stable state for a larger number of frequency bins. In the separation example of recording 12 (4 sources), shown in Fig. \ref{fig:figureRJ-4sources}, we observe that the separation quality, produced by the Reduced Likelihood Ratio Jump, is  greater to the original method, for every extracted source. Due to the accurate convergence for larger number of frequency bins, the Reduced Likelihood Ratio Jump reaches  a constant separation quality from a smaller number of iterations. In contrast, the separation quality using the original method seems to vary, depending on the permutation arising from the frequency bins that do not converge in every iteration.

Finally, in Table \ref{table4}, we present a comparison of the running times required by the 2 frameworks to produce stable separation results. The computational times of Table \ref{table4} refer to MATLAB R2013 implementations of the examined frameworks on a Pentium i7 3.4GHz PC with 8 GB RAM. As explained previously, RobustICA requires sufficiently less iterations than FastICA and in the results of Table \ref{table2} we use 3 iterations for framework 1 (RobustICA) and 15 iterations for framework 2 (FastICA). For the permutation ambiguity, both the Reduced Likelihood Ratio Jump and the Likelihood Ratio Jump required 9 iterations. We used 9 iterations that seemed to be a good choice in Fig. \ref{fig:figureRJ-4sources} since after 9 iterations on average, all sources seems to present a relevant stability in the calculated SIR values. We can see that the improvement in computational time for the examined recordings is important, with similar separation performance as shown in previous experiments. RobustICA with much less iterations requires about 1/3 of the FastICA computational time, while the Reduced Likelihood Ratio Jump can solve the permutation ambiguity in about half the time of the original LRJ. As an example, for the Recording 13, we need 19 sec with the proposed framework and 45 sec with the original one, which demonstrates the efficiency of the proposed approach.

\subsubsection{Improvement by MuSIC-RLRJ}
To evaluate the performance of the enhanced version of the framework that includes the MuSIC preprocessing step, we present the experiments of Table \ref{table5}. With the addition of beamforming, we can improve the extracted separation quality in terms of SIR. When we use only 3 iterations for RLRJ, we observe improved separation results. MuSIC works as a preprocessing step that provides RLRJ with an improved initial estimate of $\beta_j(t)$. If we allow RLRJ to perform more iterations, the algorithm overcomes the influence of beamforming and converges to performance similar to the original RLRJ.
Furthermore, in recording 12, we observe an improvement in the performance of the RLRJ for the extracted sources 3 and 4.

The MuSIC preprocessing step is not very demanding in terms of computational complexity. In Table \ref{table6}, we observe that the processing time increased by 0.5 sec from 3 to 4 sources for the MuSIC-RLRJ in contrast to 3.5 sec for RLRJ. MuSIC is actually applied only to a subset of the frequency bins that correspond to frequencies below $2500 Hz$, and its complexity increases almost proportionally to the number of sources. In contrast, RLRJ, despite being a lot faster compared to the original LRJ, still presents an $\approx O(n^2)$ complexity. As the number of sources increase, the likelihood ratio test becomes more demanding and the incorporation of geometrical information related to the auditory scene appears to increase the efficiency of our framework.

\begin{figure}
  \centering
  \begin{minipage}[b]{0.45\textwidth}
  \centering
    \includegraphics[width=1.3\textwidth, , height=\textwidth]{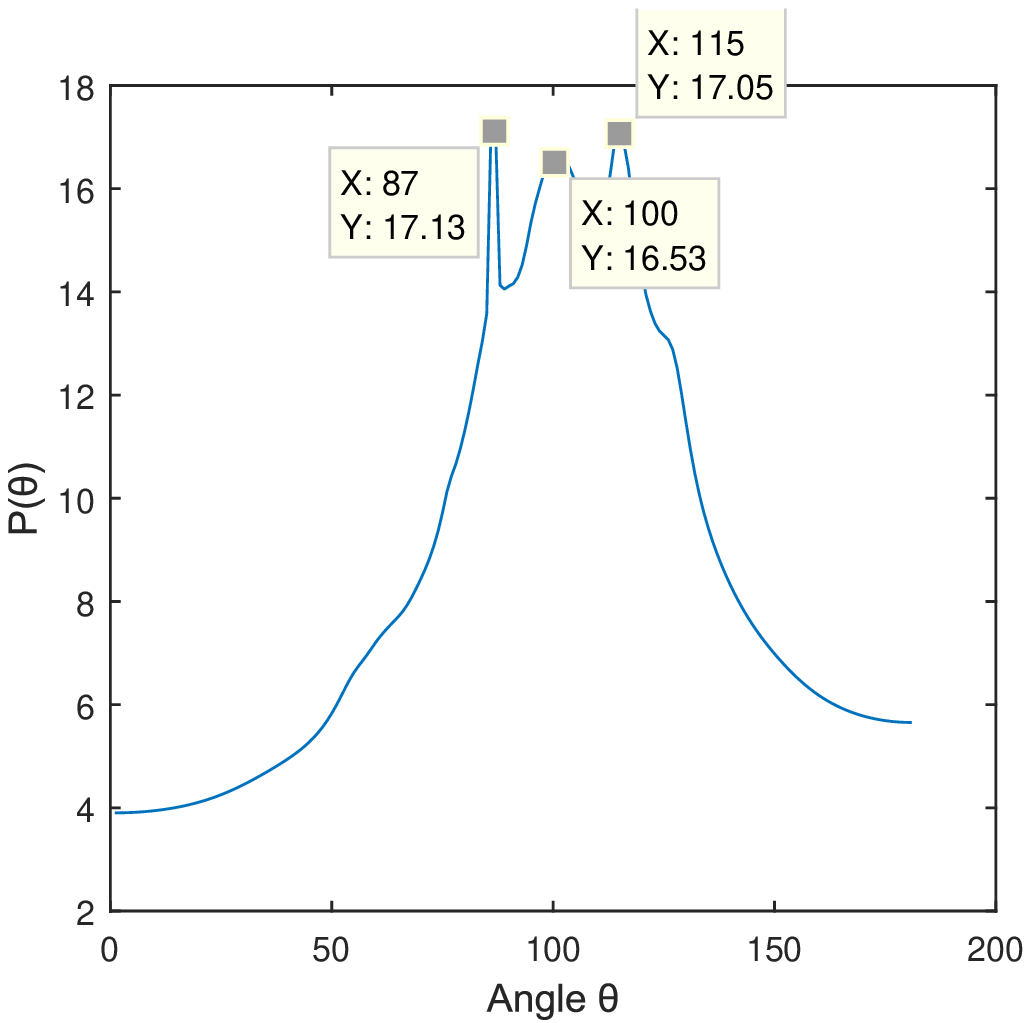}
    \caption{Direction of Arrival (DOA) estimation using MuSIC for recording 11. The DOA of the three sources are clearly visible.}
    \label{figDOA11}
  \end{minipage}
  \hfill
  \begin{minipage}[b]{0.45\textwidth}
  \centering
    \includegraphics[width=1.3\textwidth, height=\textwidth]{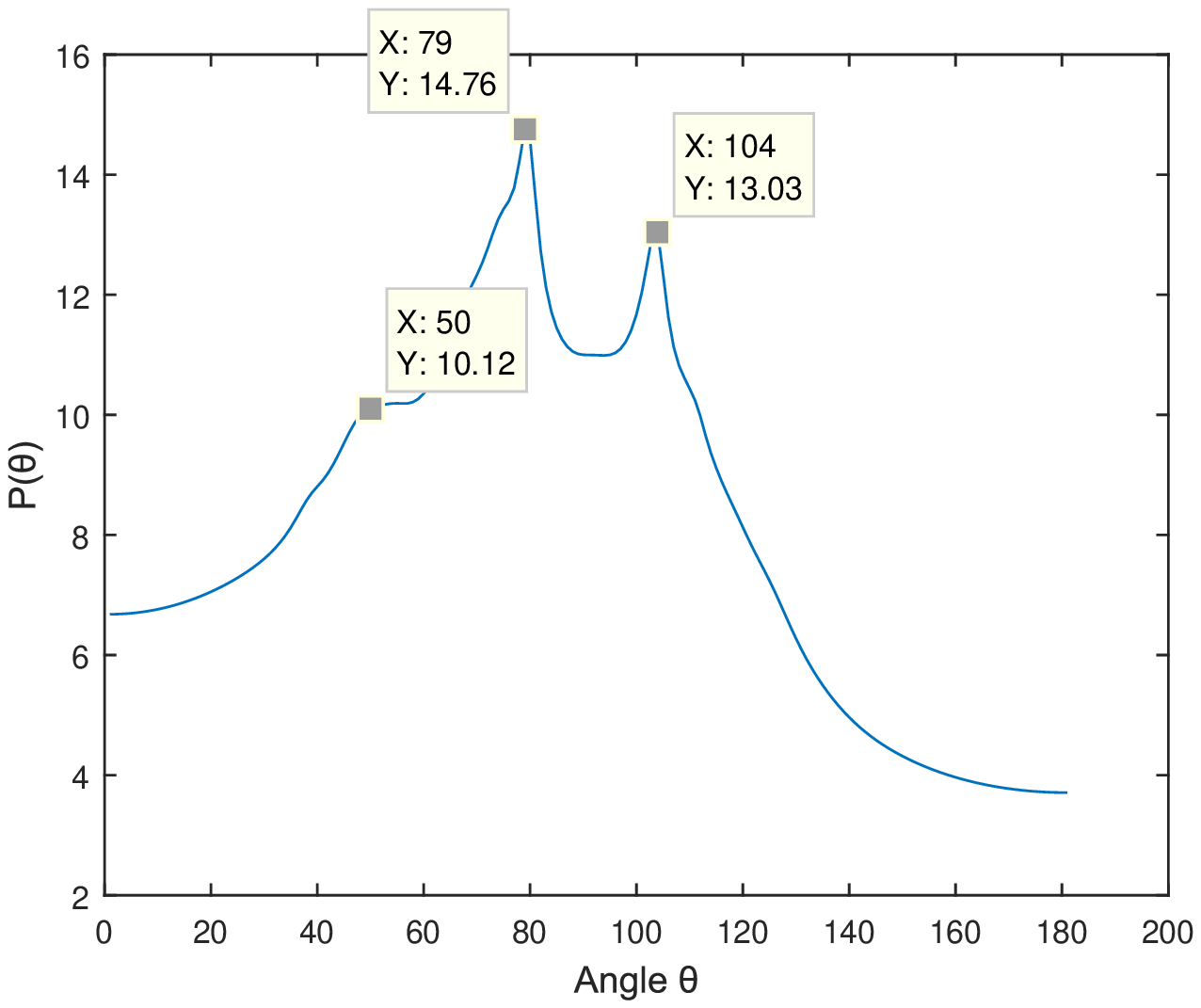}
    \caption{Direction of Arrival (DOA) estimation using MuSIC for recording 13. Due to sources' proximity, MuSIC can detect 3 out of the 4 sources.}
    \label{figDOA13}
  \end{minipage}
\end{figure}

On the other hand, beamforming is not a completely blind technique, such as ICA with LRJ. It requires information, regarding the position of sensors. In addition, it requires that the sources are angularly distant in the auditory scene, which restrains its applicability and performance. To support that we can observe Fig. \ref{figDOA11} and Fig. \ref{figDOA13}. In the DOA estimation step, all 3 observed sources come from very close angles, with angular distance below 15 degrees. Because of the source positioning in the auditory scene, the estimated bounds by MuSIC are not very distinct. As a result, the beamformer is likely to set incorrect permutations, especially between 95 and 105 degrees, where the majority of frequency bins are placed, as we can see from the increased values of $P(\theta)$. For scenarios such as this, beamforming fails to enhance the performance of RLRJ. The problem becomes more apparent for the case of recording 13. We can see that only 3 peaks can be observed. The examined recordings took place in a small real room with high reverb, and also the sources were relatively loud. These highlighted cases demonstrate the limitation of the beamforming approach, where it clearly fails to provide a good permutation estimate for the RLRJ.

\begin{table}
\renewcommand{\arraystretch}{1.3}
\caption{Efficiency of the MuSIC preprocessing step in terms of Signal-Interference Ratio (SIR) (dB) for selected recordings for 3 and 10 iterations of the Reduced Likelihood Ratio Jump (RLRJ) method. For the separation of audio sources, 3 iterations of Robust-ICA with prewhitening was used.}
\label{table5}
\centering
\begin{tabular}{cccccc}
\hline
\bfseries Rec & \bfseries Src & \multicolumn{2}{c}{3 iterations} & \multicolumn{2}{c}{10 iterations}  \\
\hline
 &  & MuSIC-RLRJ  & RLRJ  & MuSIC-RLRJ  & RLRJ  \\
 \hline
 & 1 & \textbf{0.05} &  -0.11 & -0.90 &\textbf{ -0.85} \\
9 & 2 & \textbf{2.10} &  1.95 & \textbf{1.51}  & 1.47\\
 & 3 & -6.36 &  \textbf{-6.02} & -6.41 & \textbf{-5.79}\\
\hline
 & 1 & -2.89  & \textbf{-2.65} &  -3.30 & \textbf{-2.74} \\
12 & 2 & -4.42  &  \textbf{-3.95} & -3.63 &\textbf{ -3.21} \\
 & 3 &  \textbf{3.35}  & 1.28 & \textbf{4.59} & 3.02 \\
  & 4 &  \textbf{-3.79}  & -5.04 & \textbf{-1.97} & -3.04 \\
 \hline
\end{tabular}
\end{table}

\begin{table}
\renewcommand{\arraystretch}{1.3}
\caption{Increase in running time (sec) due to the Beamforming preprocessing step. We measure the running time (sec) of the MuSIC beamforming step, followed by the Reduced Likelihood Ratio Jump (RLRJ) step for 3 and 10 iterations. MuSIC adds a delay factor to the overall running time of the algorithm.  }
\label{table6}
\centering
\begin{tabular}{cccc|cccc}
\hline
\bfseries Recording  & \multicolumn{3}{c}{3 iterations} & \multicolumn{3}{c}{10 iterations}  \\
\hline
 &   MuSIC  & RLRJ  & Total &  MuSIC  & RLRJ & Total   \\
 \hline
9 &  1.66 & 1.47 & 3.13 &  1.51  & 4.94 & 6.45\\
\hline
12  &  2.03  &  2.91 & 4.94 & 2.01 & 9.53 & 11.54 \\
 \hline
\end{tabular}
\end{table}

\subsubsection{Statistical Analysis}
{
In this section, we evaluate the statistical significance of the improvement offered by the novel methods presented here. We compare the performance of the RobustICA-Reduced Likelihood Ratio Jump (RLRJ) method (Framework 2) and the RobustICA-Beamforming-RLRJ method (Framework 3) with the original FastICA - Likelihood Ratio Jump method (Framework 1). This statistical analysis is common to machine learning literature, however, in source separation it firstly appeared recently in Simpson et al \cite{Simpson16}. Thus, we will follow their methodology in our analysis.}
{
Since our database consists of very few tracks, in order to increase the population of our results, we will segment our tracks into overlapping smaller segments and measure SIR on these small segments. We will treat the SIR score for each source as a separate sample, i.e. for a $4\times 4$ separation example segment, we will get 4 SIR scores, which will lead to 4 different samples. With this method, we manage to accummulate 140 samples, which will be sufficient to determine the statistical significance of the results. We repeat the same segmenting procedure and we acquire SIR scores for the same segments for the 3 Frameworks. Thus, we gather 3 sets of 140 measurements, one for each Framework, which will be used to evaluate the statistical significance of the improvement offered by Frameworks 2 and 3.}

{
To decide which statistical test to use, in order to perform post-hoc planned contrasts (pair-wise tests), we must look into the distribution of the three results. Empirically, we gather that the three result populations do not follow the Gaussian distribution, but a rather skewed, non-symmetric distribution. As in \cite{Simpson16}, we are prompted to use the pair-wise comparison of the models via the {\em Wilcoxon signed-rank test}, which can cater for non-Gaussian data. The Wilcoxon signed-rank test can estimate the p-values of a pair-wise comparison, i.e. the probability of the null hypothesis that the median of difference between the two models is zero, i.e. not statistically significant improvement. Great p-values demonstrate statistically non-significant improvement. Commonly in the literature, a p-value smaller than 0.05 ($p\leq 0.05$), will demonstrate statistically significant improvement. In \cite{Simpson16}, they perform a {\em Bonferroni correction}, which is essentially a division of the estimated p-value, by the number of different source-types you are comparing in the dataset. We didn't do this correction in our case, since we merged $2\times 2$, $3\times 3$ and $4\times 4$ results into a single dataset in order to increase the number of data samples. Nevertheless, as we will see, statistical significance is still detected in our experiments.}

\begin{figure}
\centering
\includegraphics[width=\textwidth]{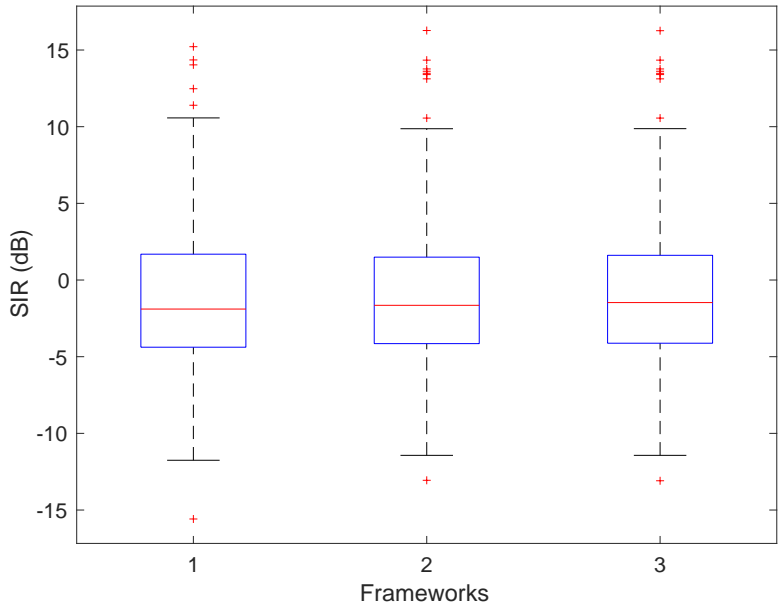}
\caption{Box-plots of the SIR performances for the three Frameworks. Medians of Frameworks 2 and 3 are slightly  higher compared to Framework 1 which is indicative of the overall offered improvement.}
\label{fig:figureboxplot}
\end{figure}

{In a similar manner to \cite{Simpson16}, Fig. \ref{fig:figureboxplot} shows box-plots for each of the Frameworks. The boxplots show the median (red), the box itself represents the inter-quartile range (IQR) and the ``whiskers'' represent $1.5\times$ IQR. Outliers (defined as data points outside the 1.5 x IQR ``whiskers'') are given as red plus signs. In Table \ref{table7}, we can see the mean and median SIR scores for the three tested Frameworks. It is clear for a first view that Frameworks 2 and 3 improve the performance on average, compared to the original Framework 1. The next step is to quantify how statistically significant is the improvement. We calculate the p-value for the two-tailed Wilcoxon signed-rank test, between Frameworks 1 and 2 ($p_{12}$) and between Frameworks 2 and 3 ($p_{23}$). The two-pair tests give $p_{12}=0.0132$ and $p_{23}=0.02$, which are well below the critical value of $p=0.05$, which denotes that the improvement offered by Frameworks 2 and 3 is statistically significant. This implies that the two presented Frameworks really improve the previous Framework 1 of Mitianoudis and Davies.}

{A final point was to see how statistically significant is the difference between Frameworks 2 and 3. We calculated the p-value for the two-tailed Wilcoxon signed-rank test between Frameworks 2 and 3 $p_{23}$. The two-pair test yields a $p_{23}= 0.5611$, which implies there is no significant difference between Frameworks 2 and 3. This implies that we may have experienced some improvement using beamforming in the previous section, however, this is not very important. This is also visible from Table \ref{table7}, where in the complete statistical test, Framework 2 has the best mean SIR value, whereas Framework 3 has the best median SIR value, giving no real winner.}

\begin{table}
\renewcommand{\arraystretch}{1.3}
\caption{ Comparing the mean and median SIR (dB) scores for the complete statistical test for the three Frameworks. It is clear that both Frameworks 2 and 3 improve performance compared to the original Framework, although it is not clear which is the best. }
\label{table7}
\centering
\begin{tabular}{c|ccc}
\hline
\bfseries Statistical  & \multicolumn{3}{c}{Framework}   \\
\bfseries Measurement  & 1&2&3   \\
\hline
 Mean& -0.9926 & \textbf{-0.5585}  & -0.5862    \\
 \hline
Median & -1.8933 &  -1.6496 &  \textbf{-1.4707}\\
 \hline
\end{tabular}
\end{table}

\section{Conclusion}
In this paper, we presented an extension of the previous work of Mitianoudis and Davies for convolutive audio source separation. First of all, the FastICA separation algorithm was replaced with the RobustICA algorithm, which improves the performance and stability of the framework. The next improvement was the proposal of a Reduced LRJ solution, in order to reduce the increased computational cost in the case of more than $2$ sensors and sources. The Reduced LRJ offers a robust solution for the more than 2 sensors-sources case, a task that is not frequently attempted in the literature. The Reduced LRJ, although a suboptimal solution, seems to achieve better separation, since it doesn't allow more source flippings than necessary. { This improvement has shown to be statistically significant in our experiments.  A MuSIC beamforming based initialisation of the sources' permutation at lower frequencies appeared to boost the performance of the RLRJ solution in several cases, however, it has shown not to be statistically significant. Not to mention that this solution can operate, provided the whole auditory scene arrangement satisfied several beamforming performance criteria.}

The new framework has been tested on a newly recorded dataset using the cost-efficient Microsoft Kinect platform with success. For future work, the authors would like to extend this framework for underdetermined recordings, i.e. dealing with real-room recordings containing more sources than sensors. This is an underdetermined problem that cannot be resolved by traditional ICA-based source separation approaches, but we can use several developments of this paper in terms of DOA estimation and time amplitude envelopes.




\bibliographystyle{splncs}
\bibliography{ref}

\begin{thebibliography}{10}

\bibitem{mitianoudisDavies}
Mitianoudis, N., Davies, M.:
\newblock Audio source separation of convolutive mixtures.
\newblock IEEE Trans. on Speech and Audio Processing \textbf{11}(5) (Sep 2003)
  489--497

\bibitem{Comon94}
Comon, P.:
\newblock Independent component analysis---a new concept?
\newblock Signal Processing \textbf{36} (1994)  287--314

\bibitem{ICAbook01}
Hyv\"{a}rinen, A., Karhunen, J., Oja, E.:
\newblock Independent Component Analysis.
\newblock John Wiley, New York (2001) 481+xxii pages.

\bibitem{ICAbook02}
Comon, P., Jutten, C.:
\newblock Handbook of Blind Source Separation: Independent Component Analysis
  and Applications.
\newblock Academic Press (2010) 856 pages.

\bibitem{seki1998blind}
Smaragdis, P.:
\newblock Blind separation of convolved mixtures in the frequency domain.
\newblock Neurocomputing \textbf{22}(1) (1998)  21--34

\bibitem{Mitiaphd}
Mitianoudis, N.:
\newblock Audio Source Separation using Independent Component Analysis.
\newblock PhD thesis, Queen Mary London (2004)

\bibitem{Parra00IEEE}
Parra, L., Spence, C.:
\newblock Convolutive blind separation of non-stationary sources.
\newblock IEEE Trans.\ on Speech and Audio Processing \textbf{8}(3) (March
  2000)  320--327

\bibitem{ikeda1998method}
Ikeda, S., Murata, N.:
\newblock A method of blind separation on temporal structure of signals.
\newblock In: ICONIP. Volume~98., Citeseer (1998)  737--742

\bibitem{Sawada03}
Sawada, H., Mukai, R., Araki, S., Makino, S.:
\newblock A robust and precise method for solving the permutation problem of
  frequency-domain blind source separation.
\newblock In: 4th Int. Symp. on Independent Component Analysis and Blind Signal
  Separation ({ICA2003}). (2003)  505--510

\bibitem{Mitia03d}
Mitianoudis, N., Davies, M.:
\newblock Using beamforming in the audio source separation problem.
\newblock In: Proc. Int.\ Symp. on {S}ignal {P}rocessing and its
  {A}pplications, Paris, France (2003)  89 -- 92

\bibitem{Mitia04e}
Mitianoudis, N., Davies, M.:
\newblock Permutation alignment for frequency domain {ICA} using subspace
  beamforming methods.
\newblock In: Proc. Int.\ Workshop on Independent Component Analysis and Source
  Separation ({ICA2004}), Granada, Spain (2004)  127--132

\bibitem{Mazur09}
Mazur, R., Mertins, A.:
\newblock An approach for solving the permutation problem of convolutive blind
  source separation based on statistical signal models.
\newblock IEEE Trans. on Audio, Speech, and Language Processing \textbf{17}(1)
  (Jan 2009)  117 -- 126

\bibitem{Sawada11}
Sawada, H., Araki, S., Makino, S.:
\newblock Underdetermined convolutive blind source separation via frequency
  bin-wise clustering and permutation alignment.
\newblock IEEE Trans. on Audio, Speech, and Language Processing \textbf{19}(3)
  (Mar 2011)  516 -- 527

\bibitem{Saito15}
Saito, S., Oishi, K., Furukawa, T.:
\newblock Convolutive blind source separation using an iterative least-squares
  algorithm for non-orthogonal approximate joint diagonalization.
\newblock IEEE Trans. on Audio, Speech, and Language Processing \textbf{23}(12)
  (Dec 2015)  2434 -- 2448

\bibitem{Sarmiento15}
Sarmiento, A., Duran-Diaz, I., Cichocki, A., Cruces, S.:
\newblock A contrast function based on generalised divergences for solving the
  permutation problem in convolved speech mixtures.
\newblock IEEE Trans. on Audio, Speech, and Language Processing \textbf{23}(11)
  (Nov 2015)  1713 -- 1726

\bibitem{Wang11}
Wang, L., Ding, H., Yin, F.:
\newblock A region-growing permutation alignment approach in frequency-domain
  blind source separation of speech mixtures.
\newblock IEEE Trans. on Audio, Speech, and Language Processing \textbf{19}(3)
  (Mar 2011)  2434 -- 2448

\bibitem{Zhang10}
Zhang, K., Chan, L.:
\newblock Convolutive blind source separation by efficient blind deconvolution
  and minimal filter distortion.
\newblock Neurocomputing \textbf{73}(13–15) (2010)  2580--2588

\bibitem{zarozoRobustICA}
Zarzoso, V., Comon, P.:
\newblock Robust independent component analysis by iterative maximization of
  the kurtosis contrast with algebraic optimal step size.
\newblock IEEE Trans. on Neural Networks \textbf{21}(2) (Feb 2010)  248--261

\bibitem{Golan17}
Markovich-Golan, S., Gannot, S., Kellermann, W.:
\newblock Combined {LCMV-TRINICON} beamforming for separating multiple speech
  sources in noisy and reverberant environments.
\newblock IEEE Trans. on Audio, Speech and Language Processing \textbf{25}(2)
  (Feb 2017)  320--332

\bibitem{Hyva99NPL}
Hyv\"arinen, A.:
\newblock The fixed-point algorithm and maximum likelihood estimation for
  independent component analysis.
\newblock Neural Processing Letters \textbf{10}(1) (1999)  1--5

\bibitem{Hyva00NN}
Hyv\"arinen, A., Oja, E.:
\newblock Independent component analysis: Algorithms and applications.
\newblock Neural Networks \textbf{13}(4-5) (2000)  411--430

\bibitem{Hyva99TNN}
Hyv\"arinen, A.:
\newblock Fast and robust fixed-point algorithms for independent component
  analysis.
\newblock IEEE Trans.\ on Neural Networks \textbf{10}(3) (1999)  626--634

\bibitem{Herbig11}
Herbig, T., Gerl, F., Minker, W., Haeb-Umbach, R.:
\newblock Adaptive systems for unsupervised speaker tracking and speech
  recognition.
\newblock Evolving Systems \textbf{2}(3) (2011)  199–--214

\bibitem{JimGriffiths}
Griffiths, L., Jim, C.:
\newblock An alternative approach to linearly constrained adaptive beamforming.
\newblock IEEE Trans.\ on Antennas and propagation \textbf{30} (1982)  27--34

\bibitem{Cheney01thelinear}
Cheney, M.:
\newblock The linear sampling method and the music algorithm.
\newblock Inverse Problems \textbf{17} (2001)  591--595

\bibitem{Moonbook}
Moon, T., Stirling, W.:
\newblock Mathematical {M}ethods and {A}lgorithms for {S}ignal {P}rocessing.
\newblock Prentice Hall, Upper Saddle River, N.J. (2000)

\bibitem{vinc}
F\'evotte, C., Gribonval, R., Vincent, E.:
\newblock {BSS EVAL Toolbox User Guide}.
\newblock Technical report, IRISA Technical Report 1706, Rennes, France, April
  2005, http://www.irisa.fr/metiss/bss eval/

\bibitem{Simpson16}
Simpson, A., Roma, G., Grais, E., Mason, R., Hummersone, C., Liutkus, A.,
  Plumbley, M.:
\newblock Evaluation of audio source separation models using hypothesis-driven
  non-parametric statistical methods.
\newblock In: 24th European Signal Processing Conference ({EUSIPCO 2016}).
  (2016)

\end{thebibliography}

\end{document}